\documentclass{emulateapj}

\shorttitle{Source Counts at 24 Microns in the FLS}
\shortauthors{Marleau et al.}

\begin{document}

\title{Extragalactic Source Counts at 24 Microns in the Spitzer First Look Survey}

\author{
Francine R. Marleau\altaffilmark{1}, 
D. Fadda\altaffilmark{1}, 
L.J. Storrie-Lombardi\altaffilmark{1},
G. Helou\altaffilmark{1},
D. Makovoz\altaffilmark{1},
D.T. Frayer\altaffilmark{1},
L. Yan\altaffilmark{1}, 
P.N. Appleton\altaffilmark{1},
L. Armus\altaffilmark{1}, 
S. Chapman\altaffilmark{2}, 
P.I. Choi\altaffilmark{1},
F. Fang\altaffilmark{1},
I. Heinrichsen\altaffilmark{1}, 
M. Im\altaffilmark{3}, 
M. Lacy\altaffilmark{1}, 
D. Shupe\altaffilmark{1},
B.T. Soifer\altaffilmark{1,2},
G. Squires\altaffilmark{1},  
J. Surace\altaffilmark{1}, 
H.I. Teplitz\altaffilmark{1},
G. Wilson\altaffilmark{1}}

\altaffiltext{1}{Spitzer Science Center, California Institute of Technology, CA 91125, USA}
\altaffiltext{2}{California Institute of Technology, CA 91125, USA}
\altaffiltext{3}{Seoul National University, Korea}

\begin{abstract}
We present the Spitzer MIPS 24~$\mu$m source counts 
in the Extragalactic First Look Survey main, verification 
and ELAIS-N1 fields. Spitzer's increased sensitivity and 
efficiency in large areal coverage over previous infrared 
telescopes, coupled with the enhanced sensitivity of the 
24 $\mu$m band to sources at intermediate redshift, 
dramatically improve the quality and statistics of number
counts in the mid-infrared. The First Look Survey 
observations cover areas of, respectively, 4.4, 0.26 
and 0.015 sq.deg. and reach 3$\sigma$ depths of 0.11, 0.08 
and 0.03 mJy. The extragalactic counts derived for each 
survey agree remarkably well. The counts can be fitted by 
a super-Euclidean power law of index $\alpha={-2.9}$ from
0.2 to 0.9 mJy, with a flattening of the counts at fluxes 
fainter than 0.2 mJy. Comparison with infrared galaxy evolution 
models reveals a peak's displacement in the 24~$\mu$m counts.
This is probably due to the detection of a new population 
of galaxies with redshift between 1 and 2, previously unseen 
in the 15~$\mu$m deep counts.
\end{abstract}


\section{Introduction}

The mid- and far-infrared regions of the electromagnetic spectrum
probe the population of starburst galaxies obscured by dust. Counting
galaxies as a function of mid and far-infrared fluxes therefore puts
new limits on the inferred geometry of the Universe. Moreover, using
counts at different wavelengths and extragalactic background light
measurements, it is possible to constrain models of galaxy evolution
(Franceschini et al. 2001; Xu et al. 2003; Lagache, Dole \& Puget
2003; Chapman et al. 2003; Chary \& Elbaz 2001). Source counts studies
have generally been done in the optical, due to the advantage of good
sensitivity and resolution of optical instruments, but optical counts are
biased by dust obscuration and the task of correcting for dust
extinction is difficult (Calzetti 1997). The optical counts reveal a
strong evolution of the population of blue galaxies as a function of
redshift to $z=1$ (Metcalfe et al. 1995; Lilly et al. 1996). At longer
near-infrared wavelengths, where the light observed samples the old
stellar population, the galaxy counts show a passive luminosity
evolution (Gardner et al. 1993; Yan et al. 1998).

Counts from mid- and far-infrared deep surveys also point to a strong
evolution.  IRAS unveiled a new population of luminous infrared
galaxies (LIGs; $L > 10^{11} L_{\odot}$) which emit the bulk of their
energy beyond 60 $\mu$m (e.g. Soifer et al. 1987). The IRAS galaxy
counts revealed an excess of faint sources compared to no-evolution
models (Hacking et al. 1987). The deep ISOCAM survey at 15~$\mu$m
discovered a population of dust-obscured galaxies at $z=0.8$ with
LIG-like luminosities (Elbaz et al. 2002; Elbaz et al. 1999; Flores et
al. 1999). The source counts derived from the deep surveys cover the
flux density range 0.05-4 mJy, showing a significant super-Euclidean
slope from 3 to 0.4 mJy and a change of slope at flux
densities fainter than 0.4 mJy. The European Large Area ISO Survey
(ELAIS; Oliver et al. 2000) has provided source counts in the flux
density range 0.45-150 mJy, linking the IRAS counts to the deep ISOCAM
counts (Gruppioni et al. 2002).

The Spitzer Space Telescope (Werner et al. 2004) was launched on
August 25, 2003 and its three cryogenically-cooled science instruments
combine to provide imaging and spectroscopy from 3 to 180~$\mu$m with
orders-of-magnitude improvements in capability over previous infrared
telescopes.  Compared to the 15~$\mu$m band whose sensitivity is
enhanced for sources in the redshift range 0.5-1.5 due to the
prominent polycyclic aromatic hydrocarbon (PAH) features, the Spitzer
24~$\mu$m band makes it possible to explore the Universe up to
$z=2.5$.  Coupled with better sensitivity and increased efficiency in
large areal coverage, the 24 $\mu$m observations with Spitzer
dramatically improve the quality and statistics of number counts in
the mid-infrared. The First Look
Survey\footnote{http://ssc.spitzer.caltech.edu/fls/} (FLS) was the
first survey undertaken by Spitzer. The purpose of this survey is to
characterize the mid-infrared sky at previously unexplored depths and
includes an extragalactic, galactic, and ecliptic component.

In this Letter, we present source counts at 24~$\mu$m in the Spitzer
FLS main and verification surveys as well as in the ELAIS-N1 region
which has deeper observations. The surveys are described in Section 2.  
We present the data reduction and flux extraction method in Section 3.  
In Section 4, we describe the sample selection and the 24~$\mu$m counts.  
Finally, in Section 5, we discuss our results and their implications.

\begin{figure}[h]
\includegraphics[width=250pt,height=250pt]{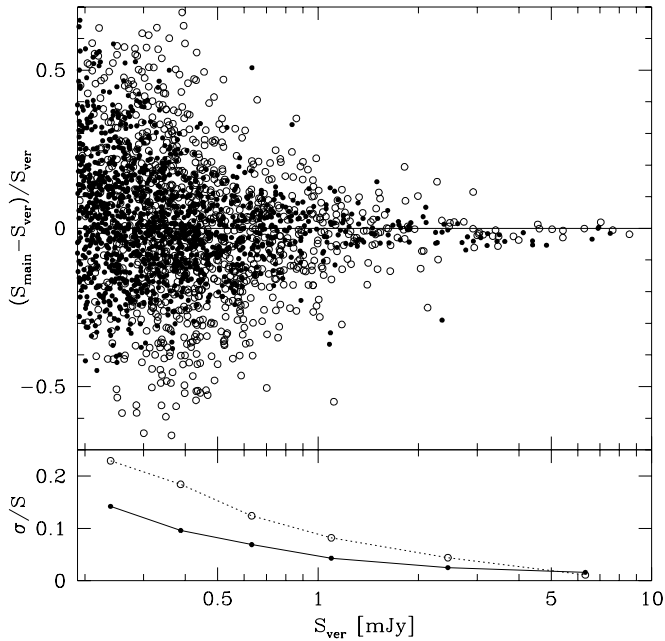}
\caption{Comparison between the StarFinder ({\it solid circles}) and
SExtractor ({\it open circles}) fluxes measured in the shallow and
deep verification region. The lower panel shows the photometric error
as a function of flux for each source extraction method. The scatter in
the difference between the shallow and deep flux measurements of faint
sources is larger for the SExtractor aperture corrected fluxes. \label{fig1}}
\end{figure}

\section{Description of the Surveys}

The extragalactic component of the Spitzer FLS covers a 4.4 square
degrees region near the ecliptic pole (17$^h$ 18$^m$
00$^s$ +59$^{\circ}$ 30$'$ 00$''$, J2000) and was observed by Spitzer
in December 2003 for a total of 62.6 hours of MIPS (Rieke et al. 2004)
and IRAC (Fazio et al. 2004) imaging. A 0.26 square degree
verification survey within the main field (17$^h$ 17$^m$
00$^s$ +59$^{\circ}$ 45$'$ 00$''$, J2000) with exposure times four
times that of the main survey was taken to allow accurate
characterization of the completeness and reliability of source
detections in the main field. In addition, deeper observations were
carried out in the ELAIS-N1 region (16$^h$ 10$^m$ 01$^s$
+54$^{\circ}$ 30$'$ 36$''$, J2000) as part of the Director's
Discretionary Time to test the confusion limit at 24~$\mu$m
(Storrie-Lombardi et al. 2004). The main characteristics of the three
surveys are given in Table~\ref{tbl-1}.

\begin{table}[h]
\begin{center}
\caption{Spitzer FLS 24$\mu$m surveys.\label{tbl-1}}
\begin{tabular}{lccccccc}
\tableline\tableline
Name &Area      &$<t_{int}>$ &Depth (3$\sigma$) &Completeness (80\%)\\
     &(sq.deg.) &(s)         &(mJy)             &(mJy)\\
\tableline
Main         &4.388 &84 &0.11 &0.23\\
Verification &0.259 &349 &0.08 &0.16\\
ELAIS-N1     &0.015 &4268 &0.03 &0.09\\
\tableline
\end{tabular}
\end{center}
\end{table}

\begin{figure}[h]
\includegraphics[width=250pt,height=250pt]{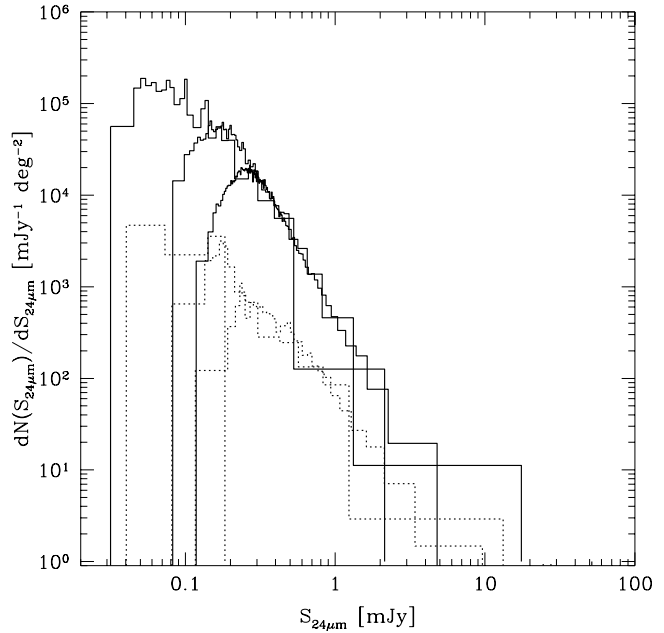}
\caption{Differential extragalactic source counts at 24~$\mu$m for the
main ({\it solid line, shallow counts}), verification ({\it solid
line, middle counts}) and ELAIS-N1 ({\it solid line, deep
counts}). The contribution of stars is displayed as separate histograms
for each survey ({\it dotted line}). \label{fig2}}
\end{figure}

\section{Data Reduction}

The FLS data were processed using the S8.9 version of the Spitzer
Science Center (SSC) pipeline (see Spitzer Observer
Manual\footnote{http://ssc.spitzer.caltech.edu/documents/som/}). Further
corrections for illumination variations and scan mirror dependent
flats were derived from the data and applied to the individual frames
(Fadda et al. 2004a). Saturated sources and bright cosmic rays can
produce a flux drop every fourth column associated with the
readouts. We identified the affected frames and applied an
additive correction to the columns with low flux.  
The frames were coadded using the SSC software
Mopex\footnote{http://ssc.spitzer.caltech.edu/postbcd/} to obtain a
mosaic with half the original pixel scale (1.27''). The projection is
done using a linear interpolation and takes into account the
distortion corrections. The initial projection utilizes bad pixel
masks produced by the pipeline. Further pixels affected by cosmic rays
are flagged using a multiframe temporal outlier detection and the
images are re-projected using these improved masks.

\begin{figure}[h]
\includegraphics[width=250pt,height=250pt]{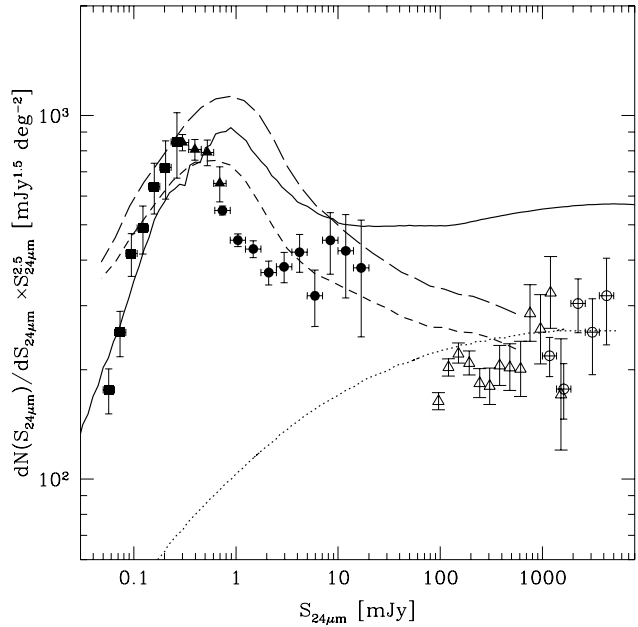}
\caption{Differential extragalactic source counts at 24~$\mu$m for the
main survey ({\it filled circles}), the verification strip ({\it
filled triangles}), and ELAIS-N1 ({\it filled squares}), normalized to
the Euclidean distribution [$N(S) \propto S^{-2.5}$]. The IRAS data
points ({\it open triangles} from Hacking \& Soifer 1991, {\it open circles} 
from Sanders et al. 2003) transformed to 24 $\mu$m are shown. 
The counts are compared to the predictions of Lagache, Dole \& Puget
(2003) ({\it solid line}), Franceschini et al. (2001) ({\it long-dash
line}) and Rodighiero et al. (2004) ({\it short-dash line}). The
no-evolution model normalized to the IRAS counts  
is shown as a {\it dotted line}. \label{fig3}}
\end{figure}

Given the resolution of MIPS at 24~$\mu$m (the PSF FWHM measures 5.9''),
typical extragalactic sources appear as point sources. For this
reason, we can perform source extraction using a PSF fitting algorithm
(StarFinder; Diolaiti et al. 2000). The StarFinder code has been
developed to analyse adaptive optics images of very crowded stellar
fields. We find that this extraction method performs remarkably well
in the case of the 24~$\mu$m images because of the good sampling
of the PSF.  Compared to the IRAF package DAOPHOT, this method performs a better
evaluation of the PSF on the image as it evaluates the background
emission over the entire frame and iteratively estimates the PSF after
each extraction (see Aloisi et al. 2001). We compared the flux
measurements obtained with StarFinder and SExtractor (Bertin \&
Arnouts 1996) in the region observed in both the shallow and
verification surveys. The scatter in the difference between the shallow
and deep flux measurements of faint sources is larger for the
SExtractor aperture corrected fluxes (see Figure~\ref{fig1}).  The
aperture used for SExtractor is 3$\times$FWHM ($\sim$70\% of
the PSF total flux).  The relative error of the flux extracted by StarFinder
is also shown in Figure~\ref{fig1}. StarFinder permits a more reliable
flux measurement as it recovers 90\% of the flux directly and deblends
sources more efficiently than SExtractor. In the case of deep fields,
deblending is essential for obtaining reliable counts. 
Point source fluxes were corrected to total fluxes using MIPS~24~$\mu$m
calibrators. The main field contains several extended sources such as
nearby galaxies and bright IRAS sources. We identified 79 extended
sources by visual inspection and measured their fluxes inside suitable
apertures after subtracting surrounding point sources.

\begin{figure}[h]
\includegraphics[width=250pt,height=250pt]{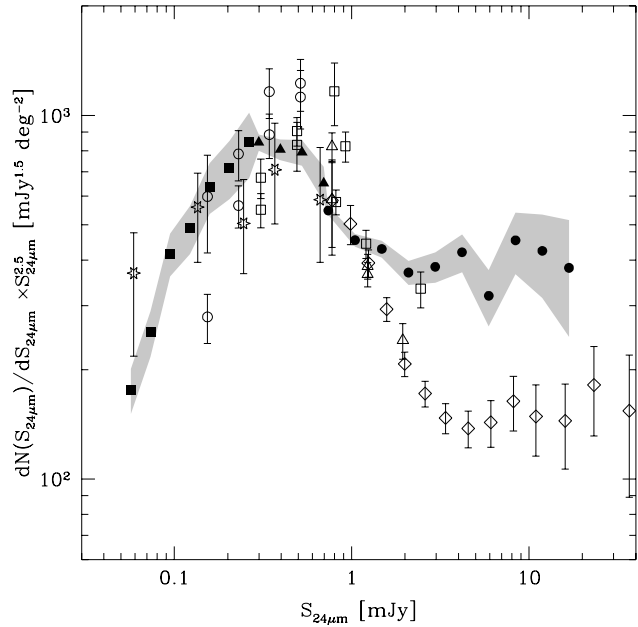}
\caption{The Spitzer FLS counts (same symbols as in Figure~\ref{fig3},
{\it shaded region} define the error bars) are compared to the ISOCAM
counts transformed to 24~$\mu$m in the HDF-N and HDF-S ({\it open
circles}; Elbaz et al. 1999), Marano Deep \& Ultra Deep fields ({\it
open squares}; Elbaz et al. 1999), in the Lockman Hole ({\it open
triangles}; Rodighiero et al. 2004), in ELAIS-S1 ({\it open diamonds};
Gruppioni et al. 2002), and through the clusters A2390, A2218 and A370
({\it open stars}; Metcalfe et al. 2003).\label{fig4}}
\end{figure}

The spacecraft astrometry (known to better than 1'') was used to coadd
the frames. A simple offset correction (of the order of 1'' in right
ascension) was applied to the final mosaic by matching sources to the
VLA counterparts (Condon et al. 2003).

\section{The 24 Micron Extragalactic Source Counts}

For each field, we considered the part of the mosaic which showed 
the most uniform coverage (number of frames coadded).
We also removed a 0.24 sq.deg. area in the main field
that contains cirrus emission which can affect
the flux measurement of sources. We counted sources in the remaining
areas (4.15, 0.26, 0.015 sq.deg.) with uniform coverage (median
of 23, 93, 139 frames) for the main, verification and ELAIS-N1
surveys, respectively.

An essential step to produce reliable counts is to discriminate stars
from extragalactic sources. The star contamination at 24~$\mu$m is
expected to be low as we are sampling the tail of the Rayleigh-Jeans
energy distribution and we are observing at high galactic latitude.
Stars are identified on the basis of their optical counterparts.  For
stars brighter than R=19, we rely on the star classification of the
Sloan Digital Sky Survey (SDSS; Hogg et al. 2004).  At fainter
magnitudes, we used the stellarity index measured with SExtractor on a
deep R-band image (Fadda et al. 2004b). This index is reliable to $R <
23$ beyond which we expect a negligible contribution of stars at
24~$\mu$m. We find 8\% stellar contribution for fluxes brighter than
0.3~mJy in the main field, 3\% in the flux range 0.2-0.3~mJy in the
verification survey, and 4\% in the 0.1-0.2~mJy range.

The completeness of the main survey was estimated using the
verification region for which both shallow and deep observations were
taken. Using the deeper verification counts, we find that the counts
in the main survey are 100\% complete to a flux of 0.28 mJy (80\%
complete at 0.23 mJy). Similarly, using the deep ELAIS-N1 counts, we
were able to measure a completeness limit of 0.21 mJy (80\% complete at 0.16 mJy)
for the verification survey. Simulations for completeness and
reliability measurements for the ELAIS-N1 field can be found in
Storrie-Lombardi et al. (2004). The completeness limits of each survey
are listed in Table~\ref{tbl-1}. The 3$\sigma$ limiting depth, also
given in Table~\ref{tbl-1}, were computed by measuring the noise
(standard deviation) in the mosaic. To compute the total limiting
flux, we assumed a Gaussian PSF with 5.9'' FWHM normalized to the flux
measured in the central pixel.

\section{Discussion and Conclusions}

The extragalactic (solid line) and stellar counts (dotted line) are
shown in Figure~\ref{fig2}. The counts from the three surveys agree
remarkably well, as shown by the Kolmogorov-Smirnov test. The probability 
that the two flux distributions (ELAIS-N1 vs.~verification and 
verification vs.~main) are not the same is less than 20\% and 15\%, 
respectively. The counts flatten at fluxes fainter than 0.2 mJy.
For fluxes between 0.1 and 0.9 mJy, all counts can be
fitted by super-Euclidean power-laws:

\begin{equation}
\frac{dN}{dS} = \left\{ \begin{array}{ll}
(724\pm97) \; S^{-2.3\pm0.6} &\ldots 0.1 < S < 0.2 \\ 
(416\pm3) \; S^{-2.9\pm0.1} &\ldots 0.2 < S < 0.9 
\end{array} 
\right.
\end{equation}

The differential counts, normalized to the expected differential
counts in an Euclidean Universe, are shown in
Figures~\ref{fig3} and \ref{fig4}. At faint fluxes ($\sim$0.1 mJy),
the average closest-neighbor separation is $\sim$17'', 
i.e. 0.12 sources per beam. This value is less than the estimated
number of sources per beam at the 5$\sigma$ confusion level 
(Takeuchi \& Ishii 2004, assuming a slope of 1.5 below 0.1 mJy).

The predictions from the infrared galaxy evolution models of
Franceschini et al. (2001; F01), Rodighiero et al. (2004), and
Lagache, Dole \& Puget (2003; LDP03) are displayed in
Figure~\ref{fig3}. The LDP03 model considers two populations of
galaxies: non-evolving normal spirals and starburst galaxies whose
luminosity density evolves with redshift. The F01 model uses 
three populations evolving differently:
non-evolving normal spirals, a fast evolving population which includes
type-II AGNs and starburst galaxies, and type-I AGNs. The type-I AGNs
evolution is based on the results from optical and X-ray observations.
The F01 model has recently been updated (Rodighiero et al. 2004) with
a different normalization to take into account the corrected counts
from the reanalysis of the ELAIS-S1 (Gruppioni et al. 2002) and the
Lockman Hole (Rodighiero et al. 2004) fields. These models fit,
besides the counts, redshift distributions and the cosmic infrared
background spectrum.

The deep ISOCAM 15~$\mu$m counts (Elbaz et al. 1999, Rodighiero et
al. 2004, Gruppioni et al. 2002, Metcalfe et al. 2003), transformed to
24~$\mu$m, are overplotted on the Spitzer FLS 24~$\mu$m counts in
Figure~\ref{fig4}. The 15 to 24~$\mu$m flux transformation was done
using template spectral energy distributions (SEDs) of three infrared
luminous galaxies with $L_{IR}=10^{12}$ (ULIG), $10^{11}$ (LIG) and
$10^{10}$ (starburst) $L_{\odot}$, taken from Chary \& Elbaz
(2001). The 24~$\mu$m/15~$\mu$m flux ratio remains fairly constant in
the redshift region 0.1-1.2 except around $z=0.5$ where the PAH
emission bands at $\sim 15 \mu$m enter the 24
$\mu$m filter while the silicate absorption feature enters the 15
$\mu$m filter. The median value we used to transform the ISOCAM counts
is 1.2. The IRAS 25~$\mu$m counts, appearing in Figure~\ref{fig3}, are
taken from Hacking \& Soifer (1991) and, for the brightest fluxes, 
are based on the revised IRAS bright galaxy sample of Sanders et
al. (2003). These counts were transformed to 24~$\mu$m counts using 
the same set of template SEDs (25~$\mu$m/24~$\mu$m flux ratio of 1.8).

The 24~$\mu$m counts confirm the existence of the rapidly evolving
dust-obscured population discovered by ISOCAM (Elbaz et al. 1999).
The main difference between the model predictions and the data, which
is reflected also in the direct comparison with the transformed ISOCAM
counts, is a shift in the turn-around of the Euclidean-normalized
counts. The peak in the 24 $\mu$m counts (at $\sim$0.2 mJy) is fainter
than what is predicted by the models. It is highly improbable that
this difference is due to problems in flux measurement and/or
calibration. The 24~$\mu$m calibration is accurate at the level of a
few percent while the uncertainty associated with flux measurements is
less than 10\% (see Section~3). The possibility of a systematic
under-evaluation of fluxes of slightly extended sources (which we
extract as point-sources with StarFinder) is also discarded by the
good agreement with aperture measurements of faint sources.


It is possible that the two bands are sampling the same population of
dust-obscured galaxies at $z\sim1$, but the 15~$\mu$m band is missing
many objects around $z=1$ and all the galaxies at $z>1.5$.  In fact,
only $\sim$60\% of the cosmic infrared background at 15~$\mu$m has
been resolved by the ISO observations (Elbaz et al. 2002).  This
hypothesis is consistent with the shift observed in the peak of the
24~$\mu$m counts to fainter fluxes compared to the peak observed in
the 15~$\mu$m counts.  Crucial information about the redshift
distribution of the 24 $\mu$m sources will come from deep
spectroscopic observations and photometric redshift studies.

\acknowledgments

We thank A.~Noriega-Crespo for useful discussions and the referee
M.~Malkan for insightful comments. We
are grateful to H.~Dole and A.~Franceschini for providing us with
their model predictions. This work is based on observations made with
the Spitzer Space Telescope, which is operated by the Jet Propulsion
Laboratory, California Institute of Technology under NASA contract
1407. Support for this work was provided by NASA through an award
issued by JPL/Caltech.

\end{document}